\documentclass[a4paper,dvips,11pt, including graphycs]{article}%
\usepackage{amsfonts}
\usepackage{graphicx}
\usepackage{amsmath}
\usepackage{amssymb}
\usepackage{latexsym}
\usepackage{color}
\usepackage{float}
\usepackage{cite}
\usepackage{makeidx}%
\providecommand{\U}[1]{\protect\rule{.1in}{.1in}}
\newcommand{\bra}{\begin{array}}
\newcommand{\era}{\end{array}}
\newcommand{\beq}{\begin{equation}}
\newcommand{\eeq}{\end{equation}}
\newcommand{\bqr}{\begin{eqnarray}}
\newcommand{\eqr}{\end{eqnarray}}

\def\BC{\bb C}
\def\_\BC{\bbi C}

\def\( {\left(}
\def\) {\right)}
\def\no2 {{\textstyle{n\over 2}}}

\begin{document}

\title{Scattering wave functions for Aharonov-Bohm-Coulomb field: \ Path integral treatment.}
\author{Nadira Boudiaf$^{1,2,\ast}$, Abdeldjalil Merdaci$^{2,3,,+}$ and Lyazid
Chetouani$^{2}$\\$^{1}$Facult\'{e} des sciences exactes, D\'{e}partement de Physique,\\Universit\'{e} Fr\`{e}res Mentouri Constantine1, 25000 Constantine, Alg\'{e}rie.\\$^{2}$Laboratoire de Physique Math\'{e}matique et Subatomique\\(LPMPS), Universit\'{e} Constantine1, 25000 Constantine, Alg\'{e}rie.\\$^{3}$Department of Physics, Faculty of Sciences,\\University of 20 ao\^{u}t 1955-Skikda, Road El-Hadaeik, B.P. 26,\\21000, Skikda, Algeria.\\$^{\ast}$boudiafnadira@gmail.com\\$^{+}$medaci69abd@gmail.com}
\maketitle

\begin{abstract}
Exact Green's functions related to Dirac particle submitted to the combination
of Aharonov-Bohm and Coulomb fields in $(2+1$) coordinate space are
analytically calculated via path integral formalism in both global and local
representations. The scattering normalized wavefunctions as well as the
corresponding continuous energy eigenvalues are extracted following this
approach. The interesting properties of the spinors are thus deduced after
symmetrization. According to the symmetric form for the Green's function, it
is shown that the equivalence with Dirac equation is undertaken with much ease.

Some particular cases are also considered.

\end{abstract}

\section{Introduction}

The $\left(  2+1\right)  $-spinor field theory has given rise to a tremendous
interest in recent years due to the existence of quantum systems that can be
effectively described by the $\left(  2+1\right)  $ Dirac equation.
Consequently, we can see that two-dimensional relativistic models of charged
fermions, related to the uniform as well as non-uniform magnetic fields can be
used to study such quantum electrodynamic effects.

The main purpose of this paper is to study the relativistic quantum
Aharonov-Bohm effect in the presence of the physical Coulomb field in $(2+1)$
coordinates space by showing how to extract the continuous energy eigenvalues
and its corresponding scattering wave functions in terms of the Green's function.

We should note that this work is a continuation of the previous paper
\cite{M21} which treated the discrete case of the Aharonov-Bohm-Coulomb field
via $\left(  2+1\right)  $path integral formalism.

To our knowledge, research on the combination of Aharonov-Bohm and Coulomb
potentials is limited except for some important works, among these latter we
quote the work done by \cite{Kha} who gave an analytical solution via
algebraic approach. However, an exact solution to the first order Dirac
equation for the Aharonov-Bohm-Coulomb field via $\left(  2+1\right)  $path
integral formalism related to scattering case, is still non-existent.

Is is well-known that the path integral formalism is becoming a working tool
of modern physics, being the easiest and most elegant means of unification. It
serves, for example, to unify the dynamics of continuous quantities such as
the position and the impulsion with that of discrete quantities such as the
spin. This\ path integral formalism gave a new breath to the research of the
analytical and exact expressions of the relativistic spinning propagators in
the presence of external fields. Generally, the advantage of this alternative
path integral formalism to the Dirac equation lies in the extraction of
existing symmetries thanks to the properties of the exact obtained Green's
function, thus the use of these symmetries allowed us to extract normalized eigenspinors.

This paper is composed of three sections and a conclusion. Section $2$ finds
exact Green's functions in $\left(  2+1\right)  $ dimensions for a combination
of the Aharonov-Bohm and Coulomb potentials in both local and global
representations. Section $3$ derives normalized scattering wave functions and
continuous energy eigenvalues based on different physical parameters and
quantum numbers and shows that the solutions provided by this research have
interesting properties and verify the first-order Dirac equation by exchanging
quantum numbers. To be more effective, it would be interesting to get the
exact and explicit results in particular cases. Therefore, section $4$ shows
that the nonrelativistic limit is undertaken with much ease according to the
symmetric form for the Green's function. The conclusion summarizes the whole
research and presents its findings.

The configuration of the field is characterized by the following features:%
\begin{equation}
A_{0}=-\tfrac{\alpha}{e\sqrt{\hat{x}^{2}+\hat{y}^{2}}},\text{ }A_{x}%
=-\tfrac{By}{x^{2}+y^{2}},\text{ }A_{y}=\tfrac{Bx}{x^{2}+y^{2}}, \label{1}%
\end{equation}
where $\tfrac{\alpha}{e}>0$ and $B$ is the flux of the magnetic field.

The propagator related to a Dirac particle in an external electromagnetic
field is the causal Green's function $S^{c}(x_{b},x_{a})$ \cite{M21,M17}%
\begin{equation}
\left(  \gamma_{\varsigma}^{\mu}\pi_{\mu}-m\right)  S^{c}(\mathbf{x}%
_{b},\mathbf{x}_{a})=-\delta^{3}(\mathbf{x}_{b}-\mathbf{x}_{a}), \label{2}%
\end{equation}
where $\pi_{\mu}=\left(  i\partial_{\mu}-eA_{\mu}\right)  ;$the subscript
$\varsigma=\pm1;$ $\eta^{\mu\nu}=diag\left(  1,-1,-1\right)  ;$ $\mathbf{x}%
^{T}\mathbf{=}\left(  x,y;t\right)  ,$

$\gamma_{\varsigma}^{\mu}=\frac{i\varsigma}{2}\epsilon^{\mu\nu\lambda}%
\gamma_{\varsigma,\nu}\gamma_{\varsigma,\lambda},\left[  \gamma_{\varsigma
}^{\mu},\gamma_{\varsigma}^{\nu}\right]  _{+}=2\eta^{\mu\nu};$ $\mu
,\nu=\overline{0,2}$

The matrices $\gamma_{\varsigma}^{\mu}$ generating the Clifford algebra are
equivalently
\begin{equation}
\gamma_{\varsigma}^{0}=i\varsigma\gamma_{\varsigma}^{1}\gamma_{\varsigma}%
^{2}=\sigma_{3},\qquad\gamma_{\varsigma}^{1}=i\varsigma\gamma_{\varsigma}%
^{0}\gamma_{\varsigma}^{2}=i\sigma_{2},\qquad\gamma_{\varsigma}^{2}%
=-i\varsigma\gamma_{\varsigma}^{0}\gamma_{\varsigma}^{1}=-i\varsigma\sigma
_{1}. \label{3}%
\end{equation}
Formally, $S^{c}(\mathbf{x}_{b},\mathbf{x}_{a})$ is the matrix element
$\langle\mathbf{x}_{b}|S^{c}|\mathbf{x}_{a}\rangle$ in the coordinate space of
the inverse Dirac operator, namely%
\begin{equation}
S^{c}=\left(  \gamma_{\varsigma}^{\mu}\pi_{\mu}-m\right)  ^{-1}, \label{4}%
\end{equation}
$S^{c}(\mathbf{x}_{b},\mathbf{x}_{a})$ is fulfilling the following equation
\begin{equation}
S^{c}(\mathbf{x}_{b},\mathbf{x}_{a})=\left(  \gamma_{\varsigma}^{\mu}\pi_{\mu
}+m\right)  S_{g}^{c}(\mathbf{x}_{b},\mathbf{x}_{a}), \label{5}%
\end{equation}
$S_{g}^{c}(\mathbf{x}_{b},\mathbf{x}_{a})$ is the global Green's function
which verifies the quadratic Dirac equation with the respect of the boundary
conditions for the $x$-space path integral
\begin{equation}
\mathbf{x}(0)=\mathbf{x}_{a},\qquad\mathbf{x}(1)=\mathbf{x}_{b}. \label{6}%
\end{equation}
It is well known that the so-called global approach is related to the square
of the Dirac equation, where the superfluous, or non-physical states, are
eliminated thereafter by activating an operator. Consequently, it finds out
that the so-called local approach is related to the first-order Dirac equation
formulation, where there is no operator and the states are all physical
without any being superfluous.

The object of the following section is to see how from the path integral
formalism, we can obtain the solution of Dirac particle interacting with the
combination of Aharonov-Bohm and Coulomb potentials by determining exactly the
corresponding Green's functions for both global and local representations.

\section{Green's functions calculations}

It is remarkable to see that the first steps of the computation of the global
description $S_{g}^{c}$ \ related to discrete treatment given in the previous
work \cite{M21} are similar to those of the continuous one. Consequently, the
Green function $S_{g}^{c}$ has the following form \cite{M21}%
\begin{align}
S_{g}^{c} &  =-\frac{i}{2}\frac{1}{\sqrt{r_{b}r_{a}}}\sum_{s=\pm1}%
\sum_{l=-\infty}^{+\infty}\frac{e^{il\left(  \varphi_{b}-\varphi_{a}\right)
}}{2\pi}\int\frac{dp_{0}}{2\pi}e^{-i\varsigma\frac{\sigma_{3}+1}{2}\varphi
_{b}}e^{+\kappa\vartheta\sigma_{2}}\mathcal{\chi}_{s}\mathcal{\chi}_{s}%
^{+}e^{-\kappa\vartheta\sigma_{2}}e^{i\varsigma\frac{\sigma_{3}+1}{2}%
\varphi_{a}}\nonumber\\
&  \times e^{-ip_{0}\left(  t_{b}-t_{a}\right)  }\left(  r_{b}\mid
r_{a}\right)  _{E_{c},l_{\mathcal{C}}\left(  s\right)  },\label{7}%
\end{align}
where%
\begin{equation}
\left(  r_{b}\mid r_{a}\right)  _{E_{c},l_{\mathcal{C}}\left(  s\right)
}=-i\tfrac{1}{\sqrt{-2E_{\mathcal{C}}}}\tfrac{\Gamma\left(  -\nu+\left\vert
l_{\mathcal{C}}\left(  s\right)  \right\vert +\frac{1}{2}\right)  }%
{\Gamma\left(  2\left\vert l_{\mathcal{C}}\left(  s\right)  \right\vert
+1\right)  }W_{\nu,\left\vert l_{\mathcal{C}}\left(  s\right)  \right\vert
}\left(  2\sqrt{-2E_{\mathcal{C}}}r_{b}\right)  M_{\nu,\left\vert
l_{\mathcal{C}}\left(  s\right)  \right\vert }\left(  2\sqrt{-2E_{\mathcal{C}%
}}r_{a}\right)  ,\label{8}%
\end{equation}
with%
\begin{align}
l_{\mathcal{C}}\left(  s\right)   &  =\sqrt{\left(  \left(  l-eB\right)
\varsigma-\frac{1}{2}\right)  ^{2}-\alpha^{2}}-\frac{\kappa s}{2},\text{
}\kappa=sign\left(  \left(  l-eB\right)  \varsigma-\frac{1}{2}\right)
.\label{9}\\
E_{\mathcal{C}} &  =\frac{1}{2}\left(  p_{0}^{2}-m^{2}\right)  ,\text{ }%
\nu=\frac{\alpha p_{0}}{\sqrt{-2E_{\mathcal{C}}}}\label{10}%
\end{align}
knowing that the spin operator $\hat{S}$ $=\sigma_{3},\sigma_{3}\mathcal{\chi
}_{s}=s\mathcal{\chi}_{s}$ ,$\sum_{s=\pm1}\mathcal{\chi}_{s}\mathcal{\chi}%
_{s}^{+}=I$ and $M_{\nu,\left\vert l_{\mathcal{C}}\left(  s\right)
\right\vert }$ , $W_{\nu,\left\vert l_{\mathcal{C}}\left(  s\right)
\right\vert }$are the Whittaker functions of the first and second kind.

We can see that the sign of $\sqrt{-2E_{\mathcal{C}}}$ involves two different
cases: the first, is related to the bound states and the corresponding
discrete energy spectrum when $E_{\mathcal{C}}$ $<0$ \cite{M21}, and the
second one, is related to the scattering states and the corresponding
continuous energy spectra when $E_{\mathcal{C}}$ $\geq0$, in what follows we
are interested to expose in details the different steps\ of \ the computation
for such case.

To evaluate the contribution of $\sqrt{-2E_{\mathcal{C}}}$ to the Green's
function, let us express (\ref{8}) as
\begin{align}
\left(  r_{b}\mid r_{a}\right)  _{E_{c},l_{\mathcal{C}}\left(  s\right)  }  &
=-i\frac{1}{\Gamma\left(  2\left\vert l_{\mathcal{C}}\right\vert +1\right)
}\int_{C}\tfrac{dz}{E_{\mathcal{C}}-\frac{z^{2}}{2}}\tfrac{\Gamma\left(
-\nu+\left\vert l_{\mathcal{C}}\right\vert +\frac{1}{2}\right)  }%
{\sqrt{-2E_{\mathcal{C}}}}\nonumber\\
&  \times W_{\nu,\left\vert l_{\mathcal{C}}\right\vert }\left(  2\sqrt
{-2E_{\mathcal{C}}}r_{b}\right)  M_{\nu,\left\vert l_{\mathcal{C}}\right\vert
}\left(  2\sqrt{-2E_{\mathcal{C}}}r_{a}\right)  , \label{11}%
\end{align}
we find it convenient to consider the following relations for the closed
contour $C,$ for more details see for example \cite{Che}
\begin{equation}
C:\left\{
\begin{array}
[c]{c}%
z=k,\quad k\in\left[  -R,+R\right] \\
z=Re^{i\phi},\quad\phi\in\left[  \pi,2\pi\right]
\end{array}
\right.  , \label{12}%
\end{equation}
and to take the limit $R\longrightarrow\infty$.

With the aid of the asymptotic behavior of the Whittaker functions the
integral over the semicircle vanishes and the equation is solved as follows:
\begin{align}
\left(  r_{b}\mid r_{a}\right)  _{E_{c}\pm i0,l_{\mathcal{C}}\left(  s\right)
}  &  =-i\frac{1}{\Gamma\left(  2\left\vert l_{\mathcal{C}}\right\vert
+1\right)  }\int_{-\infty}^{+\infty}\tfrac{dk}{E_{\mathcal{C}}\pm
i0-\frac{k^{2}}{2}}\tfrac{\Gamma\left(  -\frac{\alpha p_{0}}{\mp
ik}+\left\vert l_{\mathcal{C}}\right\vert +\frac{1}{2}\right)  }{\mp
i\left\vert k\right\vert }\nonumber\\
&  \times W_{\alpha\frac{p_{0}}{\mp i\left\vert k\right\vert },\left\vert
l_{\mathcal{C}}\right\vert }\left(  \mp2i\left\vert k\right\vert r_{b}\right)
M_{\alpha\frac{p_{0}}{\mp i\left\vert k\right\vert },\left\vert l_{\mathcal{C}%
}\right\vert }\left(  \mp2i\left\vert k\right\vert r_{a}\right)  . \label{13}%
\end{align}
We can extract the scattering wave functions from the discontinuity of the
Green's function.

The discontinuity is given by \cite{Kle}%

\begin{align}
disc\left(  r_{b}\mid r_{a}\right)  _{E_{c},l_{\mathcal{C}}\left(  s\right)  }
&  =\left(  r_{b}\mid r_{a}\right)  _{E_{c}+i0,l_{\mathcal{C}}\left(
s\right)  }-\left(  r_{b}\mid r_{a}\right)  _{E_{c}-i0,l_{\mathcal{C}}\left(
s\right)  }\nonumber\\
&  =\tfrac{2}{\Gamma^{2}\left(  2\left\vert l_{\mathcal{C}}\right\vert
+1\right)  }\int_{0}^{\infty}\frac{dk}{k}\tfrac{\left\vert \Gamma\left(
-i\frac{\alpha p_{0}}{k}+\left\vert l_{\mathcal{C}}\left(  s\right)
\right\vert +\frac{1}{2}\right)  \right\vert ^{2}}{E_{\mathcal{C}}-\frac
{k^{2}}{2}}\nonumber\\
&  \times e^{+\pi\alpha\frac{p_{0}}{k}}M_{i\alpha\frac{p_{0}}{k},\left\vert
l_{\mathcal{C}}\left(  s\right)  \right\vert }\left(  -2ikr_{b}\right)
M_{-i\alpha\frac{p_{0}}{k},\left\vert l_{\mathcal{C}}\left(  s\right)
\right\vert }\left(  2ikr_{a}\right)  ,\label{14}%
\end{align}
where we have used the following relations \cite{Abr}%

\begin{align}
M_{a,b}\left(  z\right)   &  =e^{-i\pi\left(  b+\frac{1}{2}\right)  }%
M_{-a,b}\left(  -z\right)  ,2b\neq-1,-2,-3,...\label{15}\\
M_{a,b}\left(  z\right)   &  =\Gamma\left(  2b+1\right)  e^{i\pi a}\left(
\tfrac{W_{-a,b}\left(  e^{i\pi}z\right)  }{\Gamma\left(  b-a+\frac{1}%
{2}\right)  }+e^{-i\pi\left(  b+\frac{1}{2}\right)  }\tfrac{W_{a,b}\left(
z\right)  }{\Gamma\left(  b+a+\frac{1}{2}\right)  }\right)  ,-\frac{3\pi}%
{2}<\arg z<\frac{\pi}{2} \label{16}%
\end{align}
therefore, by substituting the previous results we\ will have
\begin{align}
S_{g}^{c}\left(  \mathbf{x}_{b},\mathbf{x}_{a}\right)   &  =-\frac{i}{2}%
\tfrac{1}{\sqrt{r_{b}r_{a}}}\sum_{s=\pm1}\sum_{l=-\infty}^{+\infty}\left\{
\frac{e^{il\left(  \varphi_{b}-\varphi_{a}\right)  }}{2\pi}\int_{-\infty
}^{+\infty}\frac{dp_{0}}{2\pi}e^{-i\varsigma\frac{\sigma_{3}+1}{2}\varphi_{b}%
}e^{+\kappa\vartheta\sigma_{2}}\mathcal{\chi}_{s}\mathcal{\chi}_{s}%
^{+}e^{-\kappa\vartheta\sigma_{2}}e^{i\varsigma\frac{\sigma_{3}+1}{2}%
\varphi_{a}}\right. \nonumber\\
&  \times e^{-ip_{0}\left(  t_{b}-t_{a}\right)  }\frac{4}{\Gamma^{2}\left(
2\left\vert l_{\mathcal{C}}\right\vert +1\right)  }\int_{0}^{\infty}\frac
{dk}{k}\tfrac{\left\vert \Gamma\left(  -i\frac{\alpha p_{0}}{k}+\left\vert
l_{\mathcal{C}}\left(  s\right)  \right\vert +\frac{1}{2}\right)  \right\vert
^{2}}{p_{0}^{2}-\left(  m^{2}+k^{2}\right)  }\nonumber\\
&  \times\left.  e^{+\pi\alpha\frac{p_{0}}{k}}M_{i\alpha\frac{p_{0}}%
{k},\left\vert l_{\mathcal{C}}\left(  s\right)  \right\vert }\left(
-2ikr_{b}\right)  M_{-i\alpha\frac{p_{0}}{k},\left\vert l_{\mathcal{C}}\left(
s\right)  \right\vert }\left(  2ikr_{a}\right)  \right\}  , \label{17}%
\end{align}
with the continuous energy spectra
\begin{equation}
E_{k}=\sqrt{m^{2}+k^{2}}. \label{18}%
\end{equation}
It is easy to verify that the poles for continuous spectra are
\begin{equation}
p_{0}=\varepsilon E_{k}=\varepsilon\sqrt{m^{2}+k^{2}},\varepsilon=\pm1
\label{19}%
\end{equation}
and the property of symmetry is%
\begin{equation}
l_{\mathcal{C}}\left(  -s\right)  =\sqrt{\left(  \left(  l-eB\right)
\varsigma-\frac{1}{2}\right)  ^{2}-\alpha^{2}}+\frac{\kappa s}{2}%
=l_{\mathcal{C}}\left(  s\right)  +\kappa s. \label{20}%
\end{equation}
Let us as well use the residue theorem which gives%
\begin{equation}
\int_{-\infty}^{+\infty}f\left(  p_{0}\right)  \frac{dp_{0}}{2\pi}%
\frac{e^{-ip_{0}\left(  t_{b}-t_{a}\right)  }}{p_{0}^{2}-E_{k}^{2}}=-i%
{\textstyle\sum_{\varepsilon=\pm1}}
f\left(  \varepsilon E_{k}\right)  \frac{e^{-i\varepsilon E_{k}\left(
t_{b}-t_{a}\right)  }}{2E_{k}}\Theta\left(  \varepsilon\left(  t_{b}%
-t_{a}\right)  \right)  , \label{21}%
\end{equation}
where $\Theta(x)$ is the Heaviside function.

After a set of calculations, we obtain%
\begin{align}
S_{g}^{c}\left(  \mathbf{x}_{b},\mathbf{x}_{a}\right)   &  =-\tfrac{1}%
{\sqrt{r_{b}r_{a}}}%
{\textstyle\sum_{\varepsilon=\pm1}}
{\textstyle\sum_{s=\pm1}}
{\textstyle\sum_{l=-\infty}^{l=+\infty}}
\left\{  \frac{e^{il\left(  \varphi_{b}-\varphi_{a}\right)  }}{2\pi
}e^{-i\varsigma\frac{\sigma_{3}+1}{2}\varphi_{b}}e^{+\kappa\vartheta\sigma
_{2}}\mathcal{\chi}_{s}\mathcal{\chi}_{s}^{+}e^{-\kappa\vartheta\sigma_{2}%
}e^{i\varsigma\frac{\sigma_{3}+1}{2}\varphi_{a}}\right. \nonumber\\
&  \times\tfrac{2}{\Gamma^{2}\left(  2\kappa^{\prime}l_{\mathcal{C}}\left(
s\right)  +1\right)  }\int_{0}^{\infty}\tfrac{\left\vert \Gamma\left(
-i\frac{\alpha\varepsilon E_{k}}{k}+\kappa^{\prime}l_{\mathcal{C}}\left(
s\right)  +\frac{1}{2}\right)  \right\vert ^{2}dk}{k}\tfrac{e^{-i\varepsilon
E_{k}\left(  t_{b}-t_{a}\right)  }}{2E_{k}}\Theta\left(  \varepsilon\left(
t_{b}-t_{a}\right)  \right) \nonumber\\
&  \times\left.  e^{+\pi\alpha\frac{\varepsilon E_{k}}{k}}M_{i\alpha
\frac{\varepsilon E_{k}}{k},\kappa^{\prime}l_{\mathcal{C}}\left(  s\right)
}\left(  -2ikr_{b}\right)  M_{-i\alpha\frac{\varepsilon E_{k}}{k}%
,\kappa^{\prime}l_{\mathcal{C}}\left(  s\right)  }\left(  2ikr_{a}\right)
\right\}  ,\kappa^{\prime}=sign\left(  l_{\mathcal{C}}\left(  s\right)
\right)  . \label{22}%
\end{align}
It is easy to check the identity%

\begin{equation}%
{\textstyle\sum\limits_{s=\pm1}}
f_{s}%
{\textstyle\sum\limits_{\varepsilon=\pm1}}
g_{\varepsilon}=%
{\textstyle\sum\limits_{s=\pm1}}
f_{s}\left(  g_{s}+g_{-s}\right)  , \label{23}%
\end{equation}
and considering the mapping $s\rightarrow-s$ only for terms containing
$\Theta\left(  -s\left(  t_{b}-t_{a}\right)  \right)  ,\ $the global Green's
function related to Dirac particle is finally expressed as follows%

\begin{align}
S_{g}^{c}\left(  \mathbf{x}_{b},\mathbf{x}_{a}\right)   &  =-4%
{\textstyle\sum_{s=\pm1}}
{\textstyle\sum_{l=-\infty}^{l=+\infty}}
\frac{e^{il\left(  \varphi_{b}-\varphi_{a}\right)  }}{2\pi}\int_{0}^{\infty
}dk\tfrac{e^{-isE_{k}\left(  t_{b}-t_{a}\right)  }}{2E_{k}}e^{+\pi\alpha
\frac{sE_{k}}{k}}\Theta\left(  s\left(  t_{b}-t_{a}\right)  \right)
\nonumber\\
&  \times\left\{  e^{-i\varsigma\frac{\sigma_{3}+1}{2}\varphi_{b}}%
e^{+\kappa\vartheta\sigma_{2}}\mathcal{\chi}_{s}\mathcal{\chi}_{s}%
^{+}e^{-\kappa\vartheta\sigma_{2}}e^{i\varsigma\frac{\sigma_{3}+1}{2}%
\varphi_{a}}\right. \nonumber\\
&  \times\tfrac{\left\vert \Gamma\left(  -i\frac{\alpha sE_{k}}{k}%
+\kappa^{\prime}l_{\mathcal{C}}\left(  s\right)  +\frac{1}{2}\right)
\right\vert ^{2}}{\Gamma^{2}\left(  2\kappa^{\prime}l_{\mathcal{C}}\left(
s\right)  +1\right)  }\tfrac{M_{i\alpha\frac{sE_{k}}{k},\kappa^{\prime
}l_{\mathcal{C}}\left(  s\right)  }\left(  z_{b}\right)  }{\sqrt{z_{b}}}%
\tfrac{M_{-i\alpha\frac{sE_{k}}{k},\kappa^{\prime}l_{\mathcal{C}}\left(
s\right)  }\left(  z_{a}^{\ast}\right)  }{\sqrt{z_{a}^{\ast}}}\nonumber\\
&  +e^{-i\varsigma\frac{\sigma_{3}+1}{2}\varphi_{b}}e^{+\kappa\vartheta
\sigma_{2}}\mathcal{\chi}_{-s}\mathcal{\chi}_{-s}^{+}e^{-\kappa\vartheta
\sigma_{2}}e^{i\varsigma\frac{\sigma_{3}+1}{2}\varphi_{a}}\nonumber\\
&  \times\left.  \tfrac{\left\vert \Gamma\left(  i\frac{\alpha sE_{k}}%
{k}+\kappa^{\prime}l_{\mathcal{C}}\left(  s\right)  +\eta+\frac{1}{2}\right)
\right\vert ^{2}}{\Gamma^{2}\left(  2\kappa^{\prime}l_{\mathcal{C}}\left(
s\right)  +2\eta+1\right)  }\tfrac{M_{i\alpha\frac{sE_{k}}{k},\kappa^{\prime
}l_{\mathcal{C}}\left(  s\right)  +\eta}\left(  z_{b}\right)  }{\sqrt{z_{b}}%
}\tfrac{M_{-i\alpha\frac{sE_{k}}{k},\kappa^{\prime}l_{\mathcal{C}}\left(
s\right)  +\eta}\left(  z_{a}^{\ast}\right)  }{\sqrt{z_{a}^{\ast}}}\right\}
,\eta=\kappa^{\prime}\kappa s=\pm1. \label{24}%
\end{align}

It is very well knowing that the dynamic of the system is totally determined
by the causal Green's function $S^{c}\left(  \mathbf{x}_{b},\mathbf{x}%
_{a}\right)  $ which verifies the first order of Dirac equation given by
(\ref{2})and obtained according to (\ref{5}).

The projection operator in polar coordinates is given by
\begin{equation}
\left(  \gamma_{\varsigma}^{\mu}\pi_{\mu}+m\right)  \left(  b\right)
=\sigma_{3}\left(  i\tfrac{\partial}{\partial t_{b}}+\tfrac{\alpha}{r_{b}%
}\right)  -i\tfrac{\varsigma eB}{r_{b}}\sigma_{1}e^{i\varsigma\sigma
_{3}\varphi_{b}}-\sigma_{2}e^{i\varsigma\sigma_{3}\varphi_{b}}\tfrac{\partial
}{\partial r_{b}}+\varsigma\tfrac{1}{r_{b}}\sigma_{1}e^{i\varsigma\sigma
_{3}\varphi_{b}}\tfrac{\partial}{\partial\varphi_{b}}+m, \label{25}%
\end{equation}
we can easily check the relations
\begin{equation}
\left\{
\begin{array}
[c]{c}%
\sigma_{1}\chi_{s}=\chi_{-s}\\
\sigma_{2}\chi_{s}=is\chi_{-s}%
\end{array}
\right.  \text{ },\text{and their conjugates }\left\{
\begin{array}
[c]{c}%
\chi_{s}^{+}\sigma_{1}=\chi_{-s}^{+}\\
\chi_{s}^{+}\sigma_{2}=-is\chi_{-s}^{+}%
\end{array}
\right.  , \label{26}%
\end{equation}
for the vectors
\begin{equation}
\chi_{+1}=\left(
\begin{array}
[c]{c}%
1\\
0
\end{array}
\right)  ,\qquad\chi_{-1}=\left(
\begin{array}
[c]{c}%
0\\
1
\end{array}
\right)  . \label{27}%
\end{equation}
Then, we map (\ref{25}) and (\ref{24}) into (\ref{5}) to end up with%

\begin{align}
S^{c}\left(  \mathbf{x}_{b},\mathbf{x}_{a}\right)   &  =-4%
{\textstyle\sum_{s=\pm1}}
{\textstyle\sum_{l=-\infty}^{l=+\infty}}
\frac{e^{il\left(  \varphi_{b}-\varphi_{a}\right)  }}{2\pi}\int_{0}^{\infty
}dk\tfrac{e^{-isE_{k}\left(  t_{b}-t_{a}\right)  }}{2E_{k}}e^{+\pi\alpha
\frac{sE_{k}}{k}}\Theta\left(  s\left(  t_{b}-t_{a}\right)  \right)
\nonumber\\
&  \times e^{-i\varsigma\frac{\sigma_{3}+1}{2}\varphi_{b}}e^{+\kappa
\vartheta\sigma_{2}}\left\{  \left[  \left(  \cosh\left(  2\kappa
\vartheta\right)  E_{k}+m\right)  \mathcal{\chi}_{s}\mathcal{\chi}_{s}%
^{+}-is\pi_{s}\left(  z_{b}\right)  \mathcal{\chi}_{-s}\mathcal{\chi}_{s}%
^{+}\right]  \right. \nonumber\\
&  \times\tfrac{\left\vert \Gamma\left(  -i\frac{\alpha sE_{k}}{k}%
+\kappa^{\prime}l_{\mathcal{C}}\left(  s\right)  +\frac{1}{2}\right)
\right\vert ^{2}}{\Gamma^{2}\left(  2\kappa^{\prime}l_{\mathcal{C}}\left(
s\right)  +1\right)  }\tfrac{M_{i\alpha\frac{sE_{k}}{k},\kappa^{\prime
}l_{\mathcal{C}}\left(  s\right)  }\left(  z_{b}\right)  }{\sqrt{z_{b}}}%
\tfrac{M_{-i\alpha\frac{sE_{k}}{k},\kappa^{\prime}l_{\mathcal{C}}\left(
s\right)  }\left(  z_{a}^{\ast}\right)  }{\sqrt{z_{a}^{\ast}}}\nonumber\\
&  +\left[  \left(  -\cosh\left(  2\kappa\vartheta\right)  E_{k}+m\right)
\mathcal{\chi}_{-s}\mathcal{\chi}_{-s}^{+}+is\pi_{-s}\left(  z_{b}\right)
\mathcal{\chi}_{s}\mathcal{\chi}_{-s}^{+}\right] \nonumber\\
&  \times\left.  \tfrac{\left\vert \Gamma\left(  i\frac{\alpha sE_{k}}%
{k}+\kappa^{\prime}l_{\mathcal{C}}\left(  s\right)  +\eta+\frac{1}{2}\right)
\right\vert ^{2}}{\Gamma^{2}\left(  2\kappa^{\prime}l_{\mathcal{C}}\left(
s\right)  +2\eta+1\right)  }\tfrac{M_{i\alpha\frac{sE_{k}}{k},\kappa^{\prime
}l_{\mathcal{C}}\left(  s\right)  +\eta}\left(  z_{b}\right)  }{\sqrt{z_{b}}%
}\tfrac{M_{-i\alpha\frac{sE_{k}}{k},\kappa^{\prime}l_{\mathcal{C}}\left(
s\right)  +\eta}\left(  z_{a}^{\ast}\right)  }{\sqrt{z_{a}^{\ast}}}\right\}
e^{-\kappa\vartheta\sigma_{2}}e^{i\varsigma\frac{\sigma_{3}+1}{2}\varphi_{a}%
},\eta=\kappa^{\prime}\kappa s=\pm1, \label{28}%
\end{align}
where%
\begin{equation}
\pi_{s}\left(  z\right)  =-2ik\tfrac{\partial}{\partial z}+2iks\kappa
\tfrac{l_{\mathcal{C}}\left(  s\right)  }{z}+\kappa\alpha\tfrac{E_{k}%
}{l_{\mathcal{C}}\left(  s\right)  +\frac{\kappa s}{2}}. \label{29}%
\end{equation}
In what follows we derive normalized\ scattering wave functions and the
corresponding continuous energy eigenvalues.

\section{Continuous energy spectra and scattering states}

In what follows, we deduce the scattering wave functions and the corresponding
positive and negative continuous spectra.

Thanks to the use of the following derivative form of the confluent
hypergeometric functions \cite{Tab}%
\begin{equation}
\frac{dF}{dz}\left(  \alpha,\gamma,z\right)  =\frac{\alpha}{\gamma}F\left(
\alpha+1,\gamma+1,z\right)  , \label{30}%
\end{equation}
and the two recurrence relations \cite{Tab}%
\begin{align}
\alpha F\left(  \alpha+1,\gamma+1,z\right)   &  =\left(  \alpha-\gamma\right)
F\left(  \alpha,\gamma+1,z\right)  +\gamma F\left(  \alpha,\gamma,z\right)
,\label{31}\\
zF\left(  \alpha+1,\gamma+1,z\right)   &  =\gamma F\left(  \alpha
+1,\gamma,z\right)  -\gamma F\left(  \alpha,\gamma,z\right)  , \label{32}%
\end{align}
we easily get%

\begin{align}
\frac{dF\left(  \alpha,\gamma,z\right)  }{dz}  &  =\frac{\alpha}{\gamma
}F\left(  \alpha+1,\gamma+1,z\right)  =\frac{\alpha}{\gamma}F\left(
\alpha,\gamma,z\right)  -\frac{\alpha}{\gamma}\frac{\alpha-\gamma}{\gamma
}\frac{z}{\gamma+1}F\left(  \alpha+1,\gamma+2,z\right)  ,\label{33}\\
\frac{dF\left(  \alpha,\gamma,z\right)  }{dz}  &  =\frac{1-\gamma}{z}\left[
F\left(  \alpha,\gamma,z\right)  +\frac{\alpha-\gamma+1}{\gamma-2}\frac
{z}{\gamma-2+1}F\left(  \alpha,\gamma,z\right)  -F\left(  \alpha
-1,\gamma-2,z\right)  \right]  . \label{34}%
\end{align}
One can, immediately, verify that
\begin{align}
\frac{d}{dz}\tfrac{M_{a,b}\left(  z\right)  }{\sqrt{z}}  &  =\frac{a}%
{2b-1}\tfrac{M_{a,b}\left(  z\right)  }{\sqrt{z}}-\frac{b}{z}\tfrac
{M_{a,b}\left(  z\right)  }{\sqrt{z}}+2b\tfrac{M_{a,b-1}\left(  z\right)
}{\sqrt{z}},\label{35}\\
\frac{d}{dz}\tfrac{M_{a,b}\left(  z\right)  }{\sqrt{z}}  &  =-\frac{a}%
{2b+1}\tfrac{M_{a,b}\left(  z\right)  }{\sqrt{z}}+\frac{b}{z}\tfrac
{M_{a,b}\left(  z\right)  }{\sqrt{z}}-\tfrac{b-a+\frac{1}{2}}{1+2b}%
\tfrac{b-a+\frac{1}{2}-\left(  1+2b\right)  }{1+2b}\frac{1}{1+2b+1}%
\tfrac{M_{a,b+1}\left(  z\right)  }{\sqrt{z}}. \label{36}%
\end{align}
Consequently, we check the two interesting properties for the two values
$\eta=\pm1$
\begin{align}
&  \pi_{s}\left(  r\right)  \tfrac{M_{i\alpha\frac{sE_{k}}{k},\kappa^{\prime
}l_{\mathcal{C}}\left(  s\right)  }\left(  z\right)  }{\sqrt{z}}\nonumber\\
&  =-2ik\left(  2\kappa^{\prime}l_{\mathcal{C}}\left(  s\right)  \tfrac
{1-\eta}{2}+\tfrac{1+\eta}{2}\frac{1}{2}\tfrac{\left(  \alpha\frac{sE_{k}}%
{k}\right)  ^{2}+\left(  \kappa^{\prime}l_{\mathcal{C}}\left(  s\right)
+\frac{1}{2}\right)  ^{2}}{\left(  \kappa^{\prime}l_{\mathcal{C}}\left(
s\right)  +1\right)  \left(  2\kappa^{\prime}l_{\mathcal{C}}\left(  s\right)
+1\right)  ^{2}}\right)  \tfrac{M_{i\alpha\frac{sE_{k}}{k},\kappa^{\prime
}l_{\mathcal{C}}\left(  s\right)  +\eta}\left(  z\right)  }{\sqrt{z}},
\label{37}%
\end{align}%
\begin{align}
&  \pi_{-s}\tfrac{M_{i\alpha\frac{sE_{k}}{k},\kappa^{\prime}l_{\mathcal{C}%
}\left(  s\right)  +\eta}\left(  z\right)  }{\sqrt{z}}\nonumber\\
&  =-2ik\left(  2\left(  \kappa^{\prime}l_{\mathcal{C}}\left(  s\right)
+\eta\right)  \tfrac{1+\eta}{2}+\tfrac{1-\eta}{2}\frac{1}{2}\tfrac{\left(
\alpha\frac{sE_{k}}{k}\right)  ^{2}+\left(  \kappa^{\prime}l_{\mathcal{C}%
}\left(  s\right)  +\eta+\frac{1}{2}\right)  ^{2}}{\left(  \kappa^{\prime
}l_{\mathcal{C}}\left(  s\right)  +\eta+1\right)  \left(  2\kappa^{\prime
}l_{\mathcal{C}}\left(  s\right)  +2\eta+1\right)  ^{2}}\right)
\tfrac{M_{i\alpha\frac{sE_{k}}{k},\kappa^{\prime}l_{\mathcal{C}}\left(
s\right)  }\left(  z\right)  }{\sqrt{z}}. \label{38}%
\end{align}
We can make use of the well known formula%
\begin{align}
\Gamma\left(  z+1\right)   &  =z\Gamma\left(  z\right)  ,\Gamma\left(
z+2\right)  =z\left(  z+1\right)  \Gamma\left(  z\right)  ,\label{39}\\
\Gamma\left(  z-1\right)   &  =\frac{1}{z-1}\Gamma\left(  z\right)
,\Gamma\left(  z-2\right)  =\tfrac{1}{z-2}\tfrac{1}{z-1}\Gamma\left(
z\right)  , \label{40}%
\end{align}
or else, in more convenient forms%
\begin{align}
\Gamma\left(  z+\theta\right)   &  =\left(  z\tfrac{1+\theta}{2}+\tfrac
{1}{z-1}\tfrac{1-\theta}{2}\right)  \Gamma\left(  z\right)  ,\theta
=\pm1\label{41}\\
\Gamma\left(  z+2\theta\right)   &  =\left(  z\left(  z+1\right)
\tfrac{1+\theta}{2}+\tfrac{1}{z-2}\tfrac{1}{z-1}\tfrac{1-\theta}{2}\right)
\Gamma\left(  z\right)  ,\theta=\pm1 \label{42}%
\end{align}
by taking respectively $z=-i\frac{\alpha E_{k}}{k}+\kappa^{\prime
}l_{\mathcal{C}}\left(  s\right)  +\frac{1}{2},$ $z=-i\frac{\alpha E_{k}}%
{k}+\kappa^{\prime}l_{\mathcal{C}}\left(  s\right)  +\frac{1}{2}$ with
$\theta=\eta$, we find the following interesting properties
\begin{align}
\Gamma\left(  -i\tfrac{\alpha E_{k}}{k}+\kappa^{\prime}l_{\mathcal{C}}\left(
s\right)  +\eta+\frac{1}{2}\right)   &  =\left[  \left(  -i\tfrac{\alpha
E_{k}}{k}+\kappa^{\prime}l_{\mathcal{C}}\left(  s\right)  +\tfrac{1}%
{2}\right)  \tfrac{1+\eta}{2}+\tfrac{1}{-i\frac{\alpha E_{k}}{k}%
+\kappa^{\prime}l_{\mathcal{C}}\left(  s\right)  -\frac{1}{2}}\tfrac{1-\eta
}{2}\right] \nonumber\\
&  \times\Gamma\left(  -i\tfrac{\alpha E_{k}}{k}+\kappa^{\prime}%
l_{\mathcal{C}}\left(  s\right)  +\frac{1}{2}\right)  , \label{43}%
\end{align}%
\begin{equation}
\Gamma\left(  2\kappa^{\prime}l_{\mathcal{C}}\left(  s\right)  +1+2\eta
\right)  =\left[  2\left(  2\kappa^{\prime}l_{\mathcal{C}}\left(  s\right)
+1\right)  \left(  \kappa^{\prime}l_{\mathcal{C}}\left(  s\right)  +1\right)
\tfrac{1+\eta}{2}+\tfrac{1}{2\kappa^{\prime}l_{\mathcal{C}}\left(  s\right)
-1}\tfrac{1}{2\kappa^{\prime}l_{\mathcal{C}}\left(  s\right)  }\tfrac{1-\eta
}{2}\right]  \Gamma\left(  2\kappa^{\prime}l_{\mathcal{C}}\left(  s\right)
+1\right)  , \label{44}%
\end{equation}
thus, we get%
\begin{align}
&  \pi_{s}\left(  z\right)  G_{_{i\alpha\frac{sE_{k}}{k},\kappa^{\prime
}l_{\mathcal{C}}\left(  s\right)  }}\left(  z\right) \nonumber\\
&  =-ie^{i\eta\arctan\left(  \frac{\alpha E_{k}}{k\left(  \kappa^{\prime
}l_{\mathcal{C}}\left(  s\right)  +\frac{\eta}{2}\right)  }\right)  }%
\sqrt{\cosh\left(  2\kappa\vartheta\right)  E_{k}-m}\sqrt{\cosh\left(
2\kappa\vartheta\right)  E_{k}+m}G_{_{i\alpha\frac{sE_{k}}{k},\kappa^{\prime
}l_{\mathcal{C}}\left(  s\right)  +\eta}}\left(  z\right)  ,\label{45}\\
&  \pi_{-s}\left(  z\right)  G_{_{i\alpha\frac{sE_{k}}{k},\kappa^{\prime
}l_{\mathcal{C}}\left(  s\right)  +\eta}}\left(  z\right) \nonumber\\
&  =-ie^{-i\eta\arctan\left(  \frac{\alpha E_{k}}{k\left(  \kappa^{\prime
}l_{\mathcal{C}}\left(  s\right)  +\frac{\eta}{2}\right)  }\right)  }%
\sqrt{\cosh\left(  2\kappa\vartheta\right)  E_{k}+m}\sqrt{\cosh\left(
2\kappa\vartheta\right)  E_{k}-m}G_{_{i\alpha\frac{sE_{k}}{k},\kappa^{\prime
}l_{\mathcal{C}}\left(  s\right)  }}\left(  z\right)  , \label{46}%
\end{align}
which can be further simplified to end up with
\begin{equation}
S^{c}(\mathbf{x}_{b},\mathbf{x}_{a})=%
{\textstyle\sum\limits_{s=\pm1}}
\int_{0}^{+\infty}dk\Psi_{k.l,s}^{\varsigma}\left(  r_{b},\varphi_{b}%
;t_{b}\right)  \left(  \Psi_{k,l,s}^{\varsigma}\left(  r_{a},\varphi_{a}%
;t_{a}\right)  \right)  ^{+}\sigma_{3}s\Theta\left(  s\left(  t_{b}%
-t_{a}\right)  \right)  , \label{47}%
\end{equation}
where the scattering normalized wavefunctions are finally given by
\begin{align}
\Psi_{k,l,s}^{\varsigma}\left(  r,\varphi;t\right)   &  =\frac{2e^{il\varphi}%
}{\sqrt{2\pi}}\frac{e^{-isE_{k}t}}{\sqrt{2E_{k}}}e^{+\pi\alpha\frac{sE_{k}%
}{2k}}e^{-i\varsigma\frac{\sigma_{3}+1}{2}\varphi}e^{+\kappa\vartheta
\sigma_{2}}\nonumber\\
&  \times\left[  \sqrt{\cosh\left(  2\kappa\vartheta\right)  E_{k}%
+m}G_{_{i\alpha\frac{sE_{k}}{k},\kappa^{\prime}l_{\mathcal{C}}\left(
s\right)  }}\left(  z\right)  \mathcal{\chi}_{s}\right. \nonumber\\
&  \left.  -se^{i\eta\arctan\left(  \frac{\alpha E_{k}}{k\left(
\kappa^{\prime}l_{\mathcal{C}}\left(  s\right)  +\frac{\eta}{2}\right)
}\right)  }\sqrt{\cosh\left(  2\kappa\vartheta\right)  E_{k}-m}G_{_{i\alpha
\frac{sE_{k}}{k},\kappa^{\prime}l_{\mathcal{C}}\left(  s\right)  +\eta}%
}\left(  z\right)  \mathcal{\chi}_{-s}\right]  ,\ z=-2ikr, \label{48}%
\end{align}
with%
\begin{align}
G_{_{i\alpha\frac{sE_{k}}{k},\kappa^{\prime}l_{\mathcal{C}}\left(  s\right)
}}\left(  z\right)   &  =\tfrac{\Gamma\left(  -i\frac{\alpha E_{k}}{k}%
+\kappa^{\prime}l_{\mathcal{C}}\left(  s\right)  +\frac{1}{2}\right)  }%
{\Gamma\left(  2\kappa^{\prime}l_{\mathcal{C}}\left(  s\right)  +1\right)
}\tfrac{M_{i\alpha\frac{sE_{k}}{k},\kappa^{\prime}l_{\mathcal{C}}\left(
s\right)  }\left(  z\right)  }{\sqrt{z}},z=-2ikr,\label{49}\\
G_{_{i\alpha\frac{sE_{k}}{k},\kappa^{\prime}l_{\mathcal{C}}\left(  s\right)
+\eta}}\left(  z\right)   &  =\tfrac{\Gamma\left(  -i\frac{\alpha E_{k}}%
{k}+\kappa^{\prime}l_{\mathcal{C}}\left(  s\right)  +\eta+\frac{1}{2}\right)
}{\Gamma\left(  2\kappa^{\prime}l_{\mathcal{C}}\left(  s\right)
+2\eta+1\right)  }\tfrac{M_{i\alpha\frac{sE_{k}}{k},\kappa^{\prime
}l_{\mathcal{C}}\left(  s\right)  +\eta}\left(  z\right)  }{\sqrt{z}},z=-2ikr,
\label{50}%
\end{align}
and the corresponding positive and negative continuous energy spectra are
\begin{equation}
\mathcal{E}_{k}^{s}=s\sqrt{m^{2}+k^{2}}. \label{51}%
\end{equation}
Finally, in order to determine the phase shift, let us study the asymptotic
form of the wavefunction. The asymptotic development of confluent
hypergeometric functions is found for example in the ref \cite{Tab}%

\begin{equation}
F\left(  \lambda,\gamma;z\right)  \underset{\left\vert z\right\vert
\longrightarrow+\infty}{\longrightarrow}\frac{\Gamma\left(  \gamma\right)
}{\Gamma\left(  \lambda\right)  }e^{z}\left\vert z\right\vert ^{\lambda
-\gamma}e^{-i\frac{\pi}{2}\left(  \lambda-\gamma\right)  }+\frac{\Gamma\left(
\gamma\right)  }{\Gamma\left(  \gamma-\lambda\right)  }\left\vert z\right\vert
^{-\lambda}e^{-i\frac{\pi}{2}\lambda},z=\left\vert z\right\vert e^{-i\frac
{\pi}{2}}, \label{52}%
\end{equation}
from which we have%

\begin{equation}
M_{a,b}\left(  z\right)  \underset{\left\vert z\right\vert \longrightarrow
+\infty}{\longrightarrow}\tfrac{\Gamma\left(  1+2b\right)  }{\Gamma\left(
b-a+\frac{1}{2}\right)  }e^{\frac{z}{2}}\left\vert z\right\vert ^{-a}%
e^{i\frac{\pi}{2}a}+\tfrac{\Gamma\left(  1+2b\right)  }{\Gamma\left(
b+a+\frac{1}{2}\right)  }\left\vert z\right\vert ^{a}e^{-\frac{z}{2}%
}e^{-i\frac{\pi}{2}\left(  2b-a+1\right)  }, \label{53}%
\end{equation}
if $z=-2ikr=2kre^{-i\frac{\pi}{2}}$ the function $G_{_{i\alpha\frac{sE_{k}}%
{k},\kappa^{\prime}l_{\mathcal{C}}\left(  s\right)  }}\left(  z\right)  $
behave as%

\begin{equation}
G_{_{i\alpha\frac{sE_{k}}{k},\kappa^{\prime}l_{\mathcal{C}}\left(  s\right)
+\eta}}\left(  z\right)  \underset{r\longrightarrow+\infty}{\longrightarrow
}\tfrac{e^{-ikr}\left(  2kr\right)  ^{-i\alpha\frac{sE_{k}}{k}}e^{-\frac{\pi
}{2}\alpha\frac{sE_{k}}{k}}+e^{-2i\delta_{l},\eta}\left(  2k\right)
r^{i\alpha\frac{sE_{k}}{k}}e^{ikr}e^{-i\frac{\pi}{2}\left(  2\kappa^{\prime
}l_{\mathcal{C}}\left(  s\right)  +2\eta-i\alpha\frac{sE_{k}}{k}+1\right)  }%
}{\sqrt{2kr}}. \label{54}%
\end{equation}
Consequently, the scattering eigenspinors $\Psi_{k,l,s}^{\varsigma}\left(
r,\varphi;t\right)  $ behave as%

\begin{align}
&  \Psi_{k,l,s}^{\varsigma}\left(  r,\varphi;t\right)  \underset
{r\longrightarrow+\infty}{\longrightarrow}\frac{2e^{il\varphi}}{\sqrt{2\pi}%
}\frac{e^{-isE_{k}t}}{\sqrt{2E_{k}}}e^{-i\varsigma\frac{\sigma_{3}+1}%
{2}\varphi}e^{+\kappa\vartheta\sigma_{2}}\nonumber\\
&  \times\left[  \sqrt{\cosh\left(  2\kappa\vartheta\right)  E_{k}%
+m}e^{-i\delta_{l}}e^{-i\frac{\pi}{4}}e^{-i\pi\frac{\kappa^{\prime
}l_{\mathcal{C}}\left(  s\right)  }{2}}2\tfrac{\cos\left(  \delta
_{l}-kr-\alpha\frac{sE_{k}}{k}\ln\left(  2kr\right)  +\frac{\pi}{4}+\pi
\tfrac{\kappa^{\prime}l_{\mathcal{C}}\left(  s\right)  }{2}\right)  }%
{\sqrt{2kr}}\mathcal{\chi}_{s}\right. \nonumber\\
&  -se^{i\eta\arctan\left(  \frac{\alpha E_{k}}{k\left(  \kappa^{\prime
}l_{\mathcal{C}}\left(  s\right)  +\frac{\eta}{2}\right)  }\right)  }%
\sqrt{\cosh\left(  2\kappa\vartheta\right)  E_{k}-m}e^{-i\frac{\pi}{2}%
\kappa^{\prime}l_{\mathcal{C}}\left(  s\right)  }e^{-i\delta_{l},\eta
}e^{-i\frac{\pi}{4}}e^{-\frac{\pi}{2}\alpha\frac{sE_{k}}{k}}\nonumber\\
&  \left.  \times\tfrac{2i\sin\left(  \delta_{l},\eta-kr-\alpha\frac{sE_{k}%
}{k}\ln\left(  2kr\right)  +\frac{\pi}{2}\kappa^{\prime}l_{\mathcal{C}}\left(
s\right)  +\frac{\pi}{4}\right)  }{\sqrt{2kr}}\mathcal{\chi}_{-s}\right]
,\ z=-2ikr, \label{55}%
\end{align}
we can make sure that
\begin{align}
\Gamma\left(  \kappa^{\prime}l_{\mathcal{C}}\left(  s\right)  +\eta+\frac
{1}{2}+i\alpha\frac{sE_{k}}{k}\right)   &  =\left[  \left(  i\tfrac{\alpha
E_{k}}{k}+\kappa^{\prime}l_{\mathcal{C}}\left(  s\right)  +\tfrac{1}%
{2}\right)  \tfrac{1+\eta}{2}+\tfrac{1}{i\frac{\alpha E_{k}}{k}+\kappa
^{\prime}l_{\mathcal{C}}\left(  s\right)  -\frac{1}{2}}\tfrac{1-\eta}%
{2}\right] \nonumber\\
\times &  \left\vert \Gamma\left(  \kappa^{\prime}l_{\mathcal{C}}\left(
s\right)  +i\alpha\frac{sE_{k}}{k}+\frac{1}{2}\right)  \right\vert
e^{i\delta_{l}}, \label{56}%
\end{align}
thus, the scattering wavefunctions are finally expressed as follows%
\begin{align}
&  \Psi_{k,l,s}^{\varsigma}\left(  r,\varphi;t\right)  \underset
{r\longrightarrow+\infty}{\longrightarrow}\frac{2e^{il\varphi}}{\sqrt{2\pi}%
}\frac{e^{-isE_{k}t}}{\sqrt{E_{k}}}e^{i\delta_{l}^{\alpha}}e^{-i\delta_{AB}%
}e^{i\frac{\pi}{2}l}e^{-i\varsigma\frac{\sigma_{3}+1}{2}\varphi}%
e^{+\kappa\vartheta\sigma_{2}}\nonumber\\
&  \times\left[  \sqrt{\cosh\left(  2\kappa\vartheta\right)  E_{k}+m}%
\tfrac{\cos\left(  \delta_{AB}+\delta_{l}^{\alpha}-kr-\alpha\frac{sE_{k}}%
{k}\ln\left(  2kr\right)  -\frac{\pi l}{2}\right)  }{\sqrt{kr}}\mathcal{\chi
}_{s}\right. \nonumber\\
&  \left.  -is\sqrt{\cosh\left(  2\kappa\vartheta\right)  E_{k}-m}\tfrac
{\sin\left(  \delta_{l}^{\alpha}+\delta_{AB}+\eta\arctan\left(  \frac{\alpha
E_{k}}{k\left(  \kappa^{\prime}l_{\mathcal{C}}\left(  s\right)  +\frac{\eta
}{2}\right)  }\right)  -kr-\alpha\frac{sE_{k}}{k}\ln\left(  2kr\right)
-\frac{\pi l}{2}\right)  }{\sqrt{kr}}\mathcal{\chi}_{-s}\right]  , \label{57}%
\end{align}
where the two-phases shifts are given by%
\begin{align}
\delta_{AB}  &  =\frac{\pi}{2}\kappa^{\prime}l_{\mathcal{C}}\left(  s\right)
+\frac{\pi}{4}+\frac{\pi}{2}l,\label{58}\\
\delta_{l}^{\alpha}  &  =\arctan\left(  \Gamma\left(  \kappa^{\prime
}l_{\mathcal{C}}\left(  s\right)  +\frac{1}{2}+i\alpha\frac{sE_{k}}{k}\right)
\right)  . \label{59}%
\end{align}
In the following subsection we can make sure that\ the obtained scattering
eigenspinors are fulfilling the Dirac wave equation.

\subsection{Equivalence with Dirac equation}

The first-order Dirac equation is given by%
\begin{equation}
\left(  \gamma_{\varsigma}^{\mu}\pi_{\mu}-m\right)  \Psi_{k,l,s}^{\varsigma
}\left(  r,\varphi;t\right)  =0, \label{60}%
\end{equation}
where%

\begin{equation}
\left(  \gamma_{\varsigma}^{\mu}\pi_{\mu}-m\right)  =\sigma_{3}\left(
i\tfrac{\partial}{\partial t}+\tfrac{\alpha}{\sqrt{x^{2}+y^{2}}}\right)
-i\sigma_{2}\left(  -i\tfrac{\partial}{\partial x}+\tfrac{eBy}{x^{2}+y^{2}%
}\right)  +i\varsigma\sigma_{1}\left(  -i\tfrac{\partial}{\partial y}%
-\tfrac{eBx}{x^{2}+y^{2}}\right)  -m, \label{61}%
\end{equation}
we can make sure that the operator $\left(  \gamma_{\varsigma}^{\mu}\pi_{\mu
}-m\right)  $ in polar coordinates is
\begin{equation}
\left(  \gamma_{\varsigma}^{\mu}\pi_{\mu}-m\right)  =\sigma_{3}\left(
i\tfrac{\partial}{\partial t}+\tfrac{\alpha}{r}\right)  -i\tfrac{\varsigma
eB}{r}\sigma_{1}e^{i\varsigma\sigma_{3}\varphi}-\sigma_{2}e^{i\varsigma
\sigma_{3}\varphi}\tfrac{\partial}{\partial r}+\varsigma\tfrac{1}{r}\sigma
_{1}e^{i\varsigma\sigma_{3}\varphi}\tfrac{\partial}{\partial\varphi}-m.
\label{62}%
\end{equation}
Let us suppose that%
\begin{equation}
\left(  \gamma_{\varsigma}^{\mu}\pi_{\mu}-m\right)  \Psi_{k,l,s}^{\varsigma
}\left(  r,\varphi;t\right)  =\tfrac{2e^{il\varphi}}{\sqrt{2\pi}}%
\tfrac{e^{-isE_{k}t}}{\sqrt{2E_{k}}}e^{+\pi\alpha\frac{sE_{k}}{2k}%
}e^{-i\varsigma\frac{\sigma_{3}+1}{2}\varphi}e^{+\kappa\vartheta\sigma_{2}%
}\left(  A\mathcal{\chi}_{s}+B\mathcal{\chi}_{-s}\right)  . \label{63}%
\end{equation}
Through a straightforward calculus, we obtain
\begin{align}
A  &  =\sqrt{\cosh\left(  2\kappa\vartheta\right)  E_{k}+m}\left(
\cosh\left(  2\kappa\vartheta\right)  E_{k}-m\right)  G_{_{i\alpha\frac
{sE_{k}}{k},\kappa^{\prime}l_{\mathcal{C}}\left(  s\right)  }}\left(  z\right)
\nonumber\\
&  -se^{i\eta\arctan\left(  \frac{\alpha E_{k}}{k\left(  \kappa^{\prime
}l_{\mathcal{C}}\left(  s\right)  +\frac{\eta}{2}\right)  }\right)  }%
\sqrt{\cosh\left(  2\kappa\vartheta\right)  E_{k}-m}is\left(  \tfrac{\partial
}{\partial r}+s\kappa\tfrac{l_{\mathcal{C}}\left(  -s\right)  }{r}%
+\kappa\alpha\tfrac{E_{k}}{l_{\mathcal{C}}\left(  -s\right)  -\frac{\kappa
s}{2}}\right)  G_{_{i\alpha\frac{sE_{k}}{k},\kappa^{\prime}l_{\mathcal{C}%
}\left(  s\right)  +\eta}}\left(  z\right)  , \label{64}%
\end{align}%
\begin{align}
B  &  =-is\sqrt{\cosh\left(  2\kappa\vartheta\right)  E_{k}+m}\left(
\tfrac{\partial}{\partial r}-s\kappa\tfrac{l_{\mathcal{C}}\left(  s\right)
}{r}+\kappa\alpha\tfrac{E_{k}}{l_{\mathcal{C}}\left(  s\right)  +\frac{\kappa
s}{2}}\right)  G_{_{i\alpha\frac{sE_{k}}{k},\kappa^{\prime}l_{\mathcal{C}%
}\left(  s\right)  }}\left(  z\right) \nonumber\\
&  -se^{i\eta\arctan\left(  \frac{\alpha E_{k}}{k\left(  \kappa^{\prime
}l_{\mathcal{C}}\left(  s\right)  +\frac{\eta}{2}\right)  }\right)  }%
\sqrt{\cosh\left(  2\kappa\vartheta\right)  E_{k}-m}\left(  -\cosh\left(
2\kappa\vartheta\right)  E_{k}-m\right)  G_{_{i\alpha\frac{sE_{k}}{k}%
,\kappa^{\prime}l_{\mathcal{C}}\left(  s\right)  +\eta}}\left(  z\right)  .
\label{65}%
\end{align}
By introducing the moments $\pi_{s}\left(  z\right)  $ and $\pi_{-s}\left(
z\right)  $\ given in (\ref{29}), $A$ and $B$ take the following forms%
\begin{align}
A  &  =\sqrt{\cosh\left(  2\kappa\vartheta\right)  E_{k}+m}\left(
\cosh\left(  2\kappa\vartheta\right)  E_{k}-m\right)  G_{_{i\alpha\frac
{sE_{k}}{k},\kappa^{\prime}l_{\mathcal{C}}\left(  s\right)  }}\left(  z\right)
\nonumber\\
&  -se^{i\eta\arctan\left(  \frac{\alpha E_{k}}{k\left(  \kappa^{\prime
}l_{\mathcal{C}}\left(  s\right)  +\frac{\eta}{2}\right)  }\right)  }%
\sqrt{\cosh\left(  2\kappa\vartheta\right)  E_{k}-m}is\pi_{-s}\left(
z\right)  G_{_{i\alpha\frac{sE_{k}}{k},\kappa^{\prime}l_{\mathcal{C}}\left(
s\right)  +\eta}}\left(  z\right)  , \label{66}%
\end{align}%
\begin{align}
B  &  =-is\sqrt{\cosh\left(  2\kappa\vartheta\right)  E_{k}+m}\pi_{s}\left(
z\right)  G_{_{i\alpha\frac{sE_{k}}{k},\kappa^{\prime}l_{\mathcal{C}}\left(
s\right)  }}\left(  z\right) \nonumber\\
&  -se^{i\eta\arctan\left(  \frac{\alpha E_{k}}{k\left(  \kappa^{\prime
}l_{\mathcal{C}}\left(  s\right)  +\frac{\eta}{2}\right)  }\right)  }%
\sqrt{\cosh\left(  2\kappa\vartheta\right)  E_{k}-m}\left(  -\cosh\left(
2\kappa\vartheta\right)  E_{k}-m\right)  G_{_{i\alpha\frac{sE_{k}}{k}%
,\kappa^{\prime}l_{\mathcal{C}}\left(  s\right)  +\eta}}\left(  z\right)  .
\label{67}%
\end{align}
Then, using (\ref{45}) and (\ref{46}), we get $A=0$ and $B=0$. This confirms
that the scattering eigenspinors $\Psi_{n,l,s}^{\varsigma}\left(
r,\varphi;t\right)  $ satisfy the first order Dirac equation%
\begin{equation}
\left(  \gamma_{\varsigma}^{\mu}\pi_{\mu}-m\right)  \Psi_{n,l,s}^{\varsigma
}\left(  r,\varphi;t\right)  =0. \label{68}%
\end{equation}

\section{particular cases}

\subsection{Aharonov-Bohm potential}

At first we consider the case of the Aharonov-Bohm field \cite{Gro}, for this
we suppress the Coulomb field by setting in (\ref{48}) $\alpha\longrightarrow
0$ then $\vartheta\rightarrow0$ and $l_{\mathcal{C}}\left(  s\right)
\longrightarrow\kappa\left(  \left(  l-eB\right)  \varsigma-\frac{1+s}%
{2}\right)  ,$ we get the scattering eigenspinors%
\begin{align}
\Psi_{k,l,s}^{\varsigma}\left(  r,\varphi;t\right)   &  =\frac{2e^{il\varphi}%
}{\sqrt{2\pi}}\frac{e^{-isE_{k}t}}{\sqrt{2E_{k}}}e^{-i\varsigma\frac
{\sigma_{3}+1}{2}\varphi}\nonumber\\
&  \times\left(  \sqrt{E_{k}+m}G_{_{0,\left\vert \left(  l-eB\right)
\varsigma-\frac{1+s}{2}\right\vert }}\left(  -2ikr\right)  \mathcal{\chi}%
_{s}-s\sqrt{E_{k}-m}G_{_{0,\left\vert \left(  l-eB\right)  \varsigma
-\frac{1-s}{2}\right\vert }}\left(  -2ikr\right)  \mathcal{\chi}_{-s}\right)
, \label{69}%
\end{align}
following \cite{Abr}%
\begin{equation}
M_{0,\nu}\left(  z\right)  =2^{\nu}\Gamma\left(  \nu+1\right)  \sqrt{z}I_{\nu
}\left(  z\right)  , \label{70}%
\end{equation}
with the use of the relation between the first and second kind Bessel
functions,
\begin{equation}
I_{\nu}\left(  z\right)  =e^{-\frac{1}{2}\nu\pi i}J_{\nu}\left(  ze^{\frac
{1}{2}\pi i}\right)  ,\hspace{0in}\hspace{0in}\hspace*{0in}\hspace{0in}%
-\pi<\arg z\leqslant\tfrac{1}{2}\pi, \label{71}%
\end{equation}
and the well-known property of gamma function%
\begin{equation}
\Gamma\left(  2x\right)  =\frac{2^{2x-1}}{\sqrt{\pi}}\Gamma\left(  x\right)
\Gamma\left(  x+\tfrac{1}{2}\right)  , \label{72}%
\end{equation}
we obtain the following results
\begin{align}
G_{_{0,\left\vert \left(  l-eB\right)  \varsigma-\frac{1+s}{2}\right\vert }%
}\left(  -2ikr\right)   &  =\sqrt{\pi}e^{-\frac{1}{2}\left\vert \left(
l-eB\right)  \varsigma-\frac{1+s}{2}\right\vert \pi i}J_{\left\vert \left(
l-eB\right)  \varsigma-\frac{1+s}{2}\right\vert }\left(  kr\right)
,\label{73}\\
G_{_{0,\left\vert \left(  l-eB\right)  \varsigma-\frac{1-s}{2}\right\vert }%
}\left(  -2ikr\right)   &  =\sqrt{\pi}e^{-\frac{1}{2}\left\vert \left(
l-eB\right)  \varsigma-\frac{1-s}{2}\right\vert \pi i}J_{\left\vert \left(
l-eB\right)  \varsigma-\frac{1-s}{2}\right\vert }\left(  kr\right)  .
\label{74}%
\end{align}
thus, the two scattering eigenspinors are finally given by the following expressions%

\begin{equation}
\Psi_{k,l,s=+1}^{\varsigma}\left(  r,\varphi;t\right)  =\frac{2e^{il\varphi}%
}{\sqrt{2\pi}}\frac{e^{-iE_{k}t}}{\sqrt{2E_{k}}}\sqrt{\pi}\left(
\begin{array}
[c]{c}%
\sqrt{E_{k}+m}e^{-i\varsigma\varphi}e^{-\frac{1}{2}\left\vert \left(
l-eB\right)  \varsigma-1\right\vert \pi i}J_{\left\vert \left(  l-eB\right)
\varsigma-1\right\vert }\left(  kr\right) \\
-\sqrt{E_{k}-m}e^{-\frac{1}{2}\left\vert \left(  l-eB\right)  \varsigma
\right\vert \pi i}J_{\left\vert \left(  l-eB\right)  \varsigma\right\vert
}\left(  kr\right)
\end{array}
\right)  , \label{75}%
\end{equation}%
\begin{equation}
\Psi_{k,l,s=-1}^{\varsigma}\left(  r,\varphi;t\right)  =\frac{2e^{il\varphi}%
}{\sqrt{2\pi}}\frac{e^{+iE_{k}t}}{\sqrt{2E_{k}}}\sqrt{\pi}\left(
\begin{array}
[c]{c}%
\sqrt{E_{k}-m}e^{-i\varsigma\varphi}e^{-\frac{1}{2}\left\vert \left(
l-eB\right)  \varsigma-1\right\vert \pi i}J_{\left\vert \left(  l-eB\right)
\varsigma-1\right\vert }\left(  kr\right) \\
\sqrt{E_{k}+m}e^{-\frac{1}{2}\left\vert \left(  l-eB\right)  \varsigma
\right\vert \pi i}J_{\left\vert \left(  l-eB\right)  \varsigma\right\vert
}\left(  kr\right)
\end{array}
\right)  . \label{76}%
\end{equation}
The obtained results (\ref{75}) and (\ref{76}) are more general than those of
the ref \cite{Kha} because they are containing the phases shifts $\frac{1}%
{2}\left\vert \left(  l-eB\right)  \varsigma-1\right\vert \pi i$ and $\frac
{1}{2}\left\vert \left(  l-eB\right)  \varsigma\right\vert \pi i$ in AB-effect.

\subsection{Nonrelativistic limit}

We require the limit $\alpha\ll\left\vert l_{\mathcal{C}}\left(  s\right)
\right\vert +\frac{1}{2}$ to recover the nonrelativistic limit for Dirac
particle \cite{Gei}.

We can see that, the obtained solutions of the continuous energy spectrum are
behave as
\begin{equation}
\vartheta\left(  l\right)  \rightarrow0, \label{77}%
\end{equation}%
\begin{equation}
l_{\mathcal{C}}\left(  s\right)  \rightarrow\kappa\left(  \left(  l-eB\right)
\varsigma-\frac{1}{2}\right)  -\frac{\kappa s}{2}=\kappa\varsigma\left(
\left(  l-eB\right)  -\varsigma\frac{s+1}{2}\right)  , \label{78}%
\end{equation}

\begin{equation}
E_{k}=m\left(  1+\frac{k^{2}}{m^{2}}\right)  ^{\frac{1}{2}}\sim m+\frac{k^{2}%
}{2m}\sim m \label{79}%
\end{equation}
and then the spinor in (\ref{48}) behaves as%

\begin{equation}
\Psi_{k,l,s}^{\varsigma}\left(  r,\varphi;t\right)  \longrightarrow
\frac{2e^{il\varphi}}{\sqrt{2\pi}}e^{-is\left(  m+\frac{k^{2}}{2m}\right)
t}e^{+\pi\alpha\frac{sm}{2k}}e^{-i\varsigma\frac{s+1}{2}\varphi}%
G_{_{i\alpha\frac{sm}{k},\left\vert l-eB\right\vert }}\left(  z\right)
\mathcal{\chi}_{s}. \label{80}%
\end{equation}
Now taking $l$ $\rightarrow l+\varsigma\frac{s+1}{2}$, to end up with usual
eigenfunctions for the nonrelativistic problem
\begin{equation}
\Psi_{k,l,s=+1}^{\varsigma}\left(  r,\varphi;t\right)  \rightarrow
e^{-imt}\left(
\begin{array}
[c]{c}%
\Psi_{k,l}^{NR}\left(  r,\varphi;t\right) \\
0
\end{array}
\right)  ,\text{ } \label{81}%
\end{equation}
or%
\begin{equation}
\Psi_{k,l,s=-1}^{\varsigma}\left(  r,\varphi;t\right)  \rightarrow
e^{+imt}\left(
\begin{array}
[c]{c}%
0\\
\Psi_{k,l}^{NR}\left(  r,\varphi;t\right)
\end{array}
\right)  ,\text{ } \label{82}%
\end{equation}
where $\Psi_{k,l}^{NR}\left(  r,\varphi;t\right)  $ are the nonrelativistic
functions defined by%

\begin{equation}
\Psi_{k,l}^{NR}\left(  r,\varphi;t\right)  =\frac{2e^{il\varphi}}{\sqrt{2\pi}%
}e^{-iE^{NR}t}e^{+\pi\alpha\frac{m}{2k}}\tfrac{\Gamma\left(  -i\frac{\alpha
m}{k}+\left\vert l-eB\right\vert +\frac{1}{2}\right)  }{\Gamma\left(
2\left\vert l-eB\right\vert +1\right)  }\tfrac{M_{i\alpha\frac{m}%
{k},\left\vert l-eB\right\vert }\left(  -2ikr\right)  }{\sqrt{-2ikr}},
\label{83}%
\end{equation}
with
\begin{equation}
\tfrac{M_{i\alpha\frac{m}{k},\left\vert l-eB\right\vert }\left(  -2ikr\right)
}{\sqrt{-2ikr}}=\tfrac{e^{ikr}z^{\left\vert l-eB\right\vert +\frac{1}{2}%
}F\left(  \left\vert l-eB\right\vert -i\alpha\frac{m}{k}+\frac{1}%
{2},1+2\left\vert l-eB\right\vert ,-2ikr\right)  }{\sqrt{-2ikr}}, \label{84}%
\end{equation}
and the corresponding nonrelativistic energy is
\begin{equation}
E^{NR}=\frac{E_{k}^{2}-m^{2}}{2m}=\frac{k^{2}}{2m}. \label{85}%
\end{equation}
These results agree exactly with those given in ref\cite{Gro}.

\section{Conclusion}

\ In this paper, we have proposed for $(2+1)$-dimensional Dirac wave equation
related to a relativistic half spin particle in interaction with
Aharonov-Bohm-Coulomb field, an analytical and exact solution by means of
$(2+1)$path integral technique. We have shown following this approach that the
continuous energy eigenvalues, as well as the corresponding scattering
normalized eigenspinors are extracted from the symmetric form of the causal
Green's function.

We should note, however, that we have considered some special cases and we
have shown that the results obtained in this paper are in agreement with the literature.

\textbf{Acknowledgments}

A.Merdaci acknowledges the Deanship of Scientific Research at King Faisal
University for the financial support.

\end{document}